\documentclass[10pt,letterpaper,twocolumn]{article} 
\usepackage{ol}
\usepackage{graphicx}
\usepackage{amsmath}
\newcommand{\dvg}{d_{12}}
\newcommand{\lf}{l_{\rm F}}
\newcommand{\neff}{n}

\newcommand{\lcoh}{l_{\rm coh}}
\newcommand{\mic}{~\mu{\rm m}}
\newcommand{\rv}{{\mathbf x}}
\newcommand{\bw}{\Delta \lambda}

\begin{document}
\twocolumn[ 
\title{Tuning quadratic nonlinear photonic crystal fibers for zero
  group-velocity mismatch}
\author{Morten Bache, Hanne Nielsen, Jesper L\ae gsgaard, and Ole Bang}
\address{COM$\bullet$DTU,
Technical University of Denmark, 
  Bld. 345v, DK-2800 Lyngby, Denmark}
 \date{\today}
\begin{abstract}
  We consider an index-guiding silica photonic crystal fiber with a
  triangular air-hole structure and a poled quadratic nonlinearity.
  By tuning the pitch and the relative hole size, second-harmonic
  generation with zero group-velocity mismatch is found for any
  fundamental wavelength above 780 nm.  The phase-velocity mismatch
  has a lower limit with coherence lengths in the micron range. The
  dimensionless nonlinear parameter is inversely proportional to the
  pitch and proportional to the relative hole size. Selected cases
  show bandwidths suitable for 20 fs pulse-conversion with
  conversion efficiencies as high as 25\%/mW.
\end{abstract}
\maketitle
]

Relying on quadratic nonlinearities, second-harmonic generation (SHG)
is widely used for efficient wavelength conversion devices in order to
extend the spectral range of laser sources and to do all-optical
wavelength multiplexing. Efficient conversion from the the fundamental
to the second-harmonic (SH) mode requires a small phase mismatch
between them. Phase matching to the lowest order is typically achieved
through a quasi-phase matching (QPM) technique,\cite{fejer:1992}
whereby the group-velocity mismatch (GVM) sets the limits
to device length and bandwidth for pulsed SHG.
In conventional fibers, SHG with near-zero GVM was found for
restricted wavelengths,\cite{Kazansky:1997} while zero GVM was
predicted using mode-matching.\cite{arraf:1998} For bulk media zero
GVM was found for restricted wavelengths by spectrally noncritical
phase matching, \cite{yu:2002} and by combining non-collinear QPM with
a pulse-front tilt.
\cite{ashihara:2003} 

\begin{figure}[htb]
\centerline{\includegraphics[width=1.8cm]{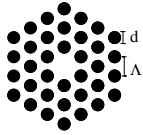}}
\caption{Sketch of a triangular structured index-guiding PCF 
with pitch $\Lambda$ and air-hole diameter $d$. }
\label{fig:pcf}
\end{figure}

Here we investigate efficient pulsed SHG in a poled silica photonic
crystal fiber (PCF), having a standard index-guiding triangular design
with a single rod defect in the center (see Fig.~\ref{fig:pcf}). The
main design parameters of the PCF are the pitch $\Lambda$ and the
relative hole size $D=d/\Lambda$.  The nonlinearity is induced,
\textit{e.g.}, by thermal poling as has recently been demonstrated in
PCFs\cite{Faccio:2001}.

When the fundamental is assumed undepleted and continuous wave, the
SHG efficiency is\cite{Kazansky:1997} $\eta\propto P_1 d_{\rm eff}^2
\lf^2 {\rm sinc}^2(\Delta \beta \lf/2)/A_{\rm ovl}$.  The
contributions to $\eta$ of nonlinear nature are the fundamental power
$P_1$, the nonlinear coefficient $d_{\rm eff}$ and the effective mode
overlap area $A_{\rm ovl}$.  Efficient SHG therefore requires a strong
mode confinement within a small core and a small effective mode
overlap area, as well as a strong nonlinear material response.  The
linear contributions to $\eta$ are the fiber length $\lf$ and the
phase mismatch $\Delta \beta=2\beta_1-\beta_2$ between the modes
($\beta_j$ are the mode propagation constants). The coherence length
of the phase mismatch to the lowest order gives the range over which
power is exchanged efficiently to the SH, but a QPM method can
compensate for this. The SHG bandwidth is determined by the width of
${\rm sinc}^2(\Delta \beta \lf/2)$, and since ultra-short pulses have
broad spectra, efficient conversion requires a large bandwidth (small
$\Delta \beta\lf$). Reducing $\lf$ does the trick, but this gives a
poor efficiency. The first order contribution to $\Delta \beta$ is
GVM, giving a temporal walk-off length $l_W$ between the modes, so we
must take $\lf\leq l_W$. When GVM is zero, $l_W\rightarrow\infty$, and
2. order dispersion takes over, giving a larger bandwidth in a longer
device.

In this Letter, we tune the phase-matching properties of SHG by
exploiting the flexibility that PCFs offer in designing the dispersion
properties. \cite{Ferrando:2000} Previous
investigations\cite{monro:2001} of SHG in PCFs considered the scalar
case and found large bandwidths and strong modal overlaps for selected
parameter values. Instead, we perform a detailed vectorial analysis
over a continuous parameter space, and show zero GVM for any
fundamental wavelength $\lambda_1>780$~nm by merely adjusting the
pitch and the relative hole size. This is a much simpler way of
removing GVM compared to previous
methods,\cite{arraf:1998,yu:2002,ashihara:2003} and promises very
large bandwidths due to its flexibility. The nonlinear properties of
the SHG fiber design are also discussed.

We first focus on the linear dispersion. A fiber mode can be described
by an effective index $\neff=c/v_{\rm ph}$, \textit{i.e.}, the ratio
of the speed of light $c$ to the phase velocity of the mode $v_{\rm
  ph}=\omega/\beta$.  The dispersive character of $\beta$ gives a
phase-velocity mismatch between the modes with frequencies $\omega_1$
(fundamental) and $\omega_2=2\omega_1$ (SH), which we classify through
the \textit{index mismatch} $\Delta n=c[1/v_{\rm
  ph}(\omega_1)-1/v_{\rm ph}(\omega_2)]
=c[\beta_1/\omega_1-\beta_2/\omega_2]$. For SHG the index mismatch is
related to the lowest order phase mismatch as $\Delta \beta=4\pi\Delta
n/\lambda_1$.  The mode group velocity is instead defined as $1/v_{\rm
  g}=\partial_\omega\beta$, $\partial_\omega\equiv
\frac{\partial}{\partial \omega}$, giving a GVM (walk-off) parameter
$d_{12}=[1/v_{\rm g}(\omega_1)-1/v_{\rm g}(\omega_2)]$.

We calculated the dispersion with the MIT Photonic-Bands (MPB)
package.\cite{johnson:2001} Each unit cell contained $n_{\rm
  C}^2=32^2$ grid points, and the super cell contained $n_{\rm
  SC}^2=5^2$ unit cells. The fundamental mode frequency and
group velocity were first calculated, followed by iterative
calculations of the SH until
$|\omega_2-2\omega_1|<10^{-8}$. Subsequently, a
perturbative approach\cite{laegsgaard:2003} was used to introduce
chromatic dispersion, allowing us to calculate data once over a large
$(D,\beta_1)$ parameter space for $\Lambda$ unity, and perturbatively
calculate the changes as $\Lambda$ was varied.

\begin{figure}[htb]
\centerline{
\includegraphics[width=8.4cm]{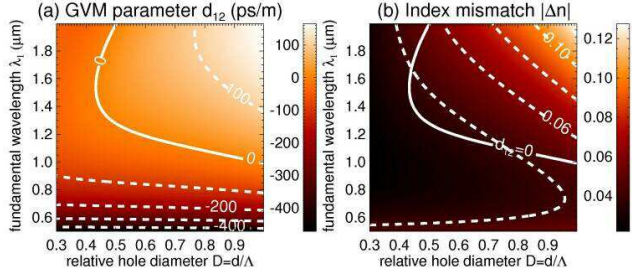}} 
\caption{Color maps of (a) $\dvg$
  and (b) $|\Delta n|$ as function of $D$ and $\lambda_1$, keeping the
  pitch $\Lambda=1.6 ~\mu$m fixed. The solid contour in (a) shows
  $\dvg=0$ and it is repeated in (b). }
\label{Fig:contours}
\end{figure}

\begin{figure}[htb]
  \centerline{ \includegraphics[width=7.5cm]{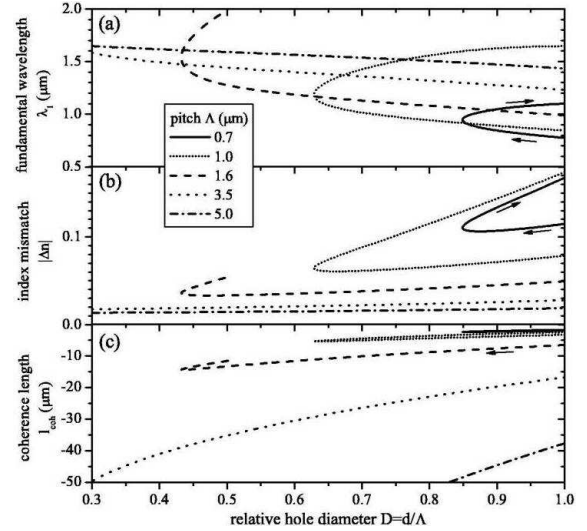}}
\caption{Extracted zero GVM contours showing (a) $\lambda_1$, (b)
  $|\Delta n|$ and (c) $\lcoh$. The curves in (b)-(c) for
  $\Lambda=1.6~\mu$m abruptly stop because $\lambda_1\in
  [0.5,2]~\mu$m. The arrows indicate the contour-direction for
  increasing $\lambda_1$.
}
\label{Fig:lambda-phase}
\end{figure}

Figure~\ref{Fig:contours} shows the GVM and index mismatch in the
$(D,\lambda_1)$ parameter space, keeping the pitch fixed at
$\Lambda=1.6~\mu$m. Along the the solid contour $d_{12}=0$: thus, zero
GVM is possible for any $\lambda_1>1\mic$ by choosing a proper $D$.
Fig.~\ref{Fig:lambda-phase}(a) underlines that this is a general
trend: there the zero-GVM contour is shown for selected pitches, and
we found $\dvg=0$ can be achieved for any $\lambda_1>0.78\mic$. Note
that in Fig.~\ref{Fig:contours}(b) $|\Delta n|>0$, even with non-zero
GVM, so the index mismatch cannot be zero. This holds also for other
$\Lambda$ values, so efficient SHG will require additional
phase-matching such as QPM.

Figure~\ref{Fig:lambda-phase}(b) shows the values of the index
mismatch as the zero-GVM contour is traversed.  For
$\Lambda=0.70,~1.0$ and $1.6~\mu$m a cusp appears around
$\lambda_1\simeq \Lambda$, after which $|\Delta n|$
increases with $\lambda_1$. This is because the fundamental mode is no
longer well-confined in the core while the SH, having a smaller
wavelength, is still well confined.  Conversely, for the considered
wavelengths the modes are always well confined for
$\Lambda=3.5,~5.0~\mu$m, explaining why a small $|\Delta n|$ is
observed there. The coherence length $\lcoh=\pi/\Delta
\beta=\lambda_1/(4\Delta n)$ is shown for completeness in
Fig.~\ref{Fig:lambda-phase}(c), giving typical values in the micron
range.

\begin{figure}[htb]
\centerline{
\includegraphics[width=7.cm]{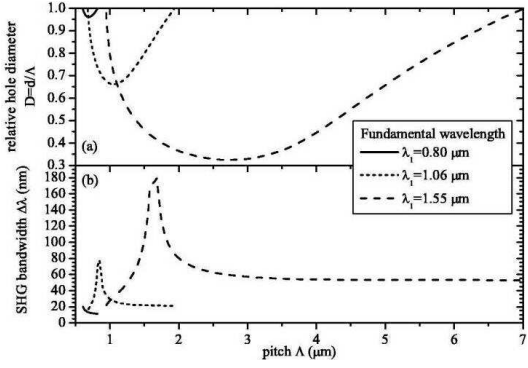}} 
\caption{Zero GVM contours vs. $\Lambda$ with fixed
  $\lambda_1$s, showing (a) $D$ as well as (b) $\bw$ for a $\lf=10$~cm
  device. }
\label{Fig:lambda-1-155}
\end{figure}

Let us focus on the telecom, Nd:YAG and Ti:Sapphire operating
wavelengths ($\lambda_1=1.55\mic,~1.06\mic$ and $0.80\mic$,
respectively.) In Fig.~\ref{Fig:lambda-1-155}(a) we then show the
$D$-value required to get zero GVM as $\Lambda$ is changed. For
$\lambda_1=0.80~\mu$m zero GVM requires very large $D$-values,
\textit{e.g.}, $D=0.96$ for $\Lambda=0.70$. For such $D$-values
deviations from the ideal circular holes must be expected, which might
influence the results presented here. However, to our knowledge it is
the first time that zero GVM has been demonstrated for SHG in any
material for this wavelength. For $\lambda_1=1.06~\mu$m the lowest
required $D$-values are in a range where the ideal round holes should
be preserved. The curves stop for larger $\Lambda$ because it is no
longer possible to get $\dvg=0$ (it would require $D>1$, which is
unphysical.)  For $\lambda_1=1.55~\mu$m a wide range of possibilities
are offered, but it is advantageous to have $\Lambda<2\mic$ because
$D$ increases, which leads to higher intensities due to stronger mode
confinement in a smaller core diameter $d_{\rm c}=\Lambda(2-D)$. In
Fig.~\ref{Fig:lambda-1-155} we calculate the SHG bandwidth $\bw$ of
the $\mathrm{ sinc}^2(\Delta \beta \lf/2)$ term by
expanding\cite{fejer:1992,ashihara:2003} $\Delta
\beta(\lambda_1+\bw)=\sum_{m}(m!)^{-1}\bw^m\partial_\lambda^m \Delta
\beta$, and assuming that a QPM grating compensates the
$m=0$ term (as is routinely done in conventional fibers
\cite{Kazansky:1997}). Since $\dvg=0$, the 2. order dispersion
dominates yielding very large bandwidths. Moreover, because $\dvg=0$
the bandwidth scales as $\bw \propto \lf^{-1/2}$ (instead of $\bw
\propto \lf^{-1}$ for $\dvg\neq 0$), and thus a longer device can be
created without loosing too much bandwidth. Note also in
Fig.~\ref{Fig:contours}(a) the turn of the zero-GVM contour around
$D=0.43$ and $\lambda_1=1.6\mic$, implying that the $m=2$ term
vanishes, giving an increasing bandwidth as observed in
Fig.~\ref{Fig:lambda-1-155}(b). It is still unclear to what extend the
exact location of this turning point is influenced by parameter
uncertainties.

Using the reductive perturbation method \cite{kodama:1987}, the
dimensionless nonlinear equations for SHG are
\begin{eqnarray}
  \label{eq:FH}
  (\partial_z
  -i \tilde D_1\partial_{t}^2 )u_1
= i\sigma u_1^*  u_2 e^{-i\Delta \beta zl_{\rm F}},
\\
  (\partial_z-\tilde{d}_{12}\partial_t
  -i \tilde D_2\partial_{t}^2 )u_2
= i\sigma u_1^2/2 e^{i\Delta \beta zl_{\rm F}},
  \label{eq:SH}
\end{eqnarray}
We have assumed a weak nonlinearity [so ${\mathbf E}_j({\mathbf r})
=A_j(z,t){\mathbf e}_j(\rv)e^{i(\beta_j z-\omega_jt)}+{\rm c.c.}$,
where $\rv=(x,y)$,] uniform over the silica part of the super cell,
and weak transverse variations in the refractive index so $\nabla
\times (\nabla \times {\mathbf E})=-\nabla^2{\mathbf E}$. The
``retarded'' coordinate $z$ [traveling with velocity $v_{\rm
  g}(\omega_1)$] is normalized to $l_{\rm F}$, $t$ to the input pulse
length $\tau$, $\tilde d_{12}=d_{12}\lf/\tau$ and $\tilde
D_j=\lf/(2\tau^2) \partial_{\omega}^2 \beta_j $. We define
$A_j=u_j(2\Lambda^2\lf/ N_jn_ja_j c \tau)^{1/2}$ so $N_j(z)=\int
dt|u_j(z,t)|^2$ is the photon number in the $j$th wave.  The
dimensionless nonlinear parameter is $ \sigma =\rho l_{\rm F} (2\hbar
\omega_1^2 \omega_2/n_1^2n_2\varepsilon_0 c^3 \tau)^{1/2}$,
where 
$\rho=(a_1^2a_2)^{-1/2}|\int d\rv \tilde{\mathbf e}_1^*(\rv)\cdot
\tilde{\chi}^{(2)}(\rv):\tilde {\mathbf e}_2(\rv)\tilde {\mathbf
  e}_1^*(\rv)| $, $\tilde{\chi}^{(2)}(\rv)$ is the Fourier transform
of the quadratic nonlinear dielectric tensor, $\tilde {\mathbf
  e}_j(\rv)={\mathbf e}_j(\rv)(\varepsilon_0 \Lambda^2 \lf/\hbar
\omega_j N_j)^{1/2}$ the dimensionless transverse modes from MPB and
$a_j=\int d\rv |\tilde {\mathbf e}_j(\rv)|^2 $ the mode areas.
Integrating over $z$ in Eq.~(\ref{eq:SH}) gives $\eta=
P_1\rho^2\lf^2\mathrm{sinc}^2(\Delta \beta \lf/2)
2\omega_1^2/n_1^2n_2\varepsilon_0 c^3$. Thus, $\rho^2$ is the
vectorial equivalent of $d_{\rm eff}^2/A_{\rm ovl}$ of
Ref.~\citeonline{Kazansky:1997}.

\begin{figure}[tb]
  \centerline{ \includegraphics[width=8cm]{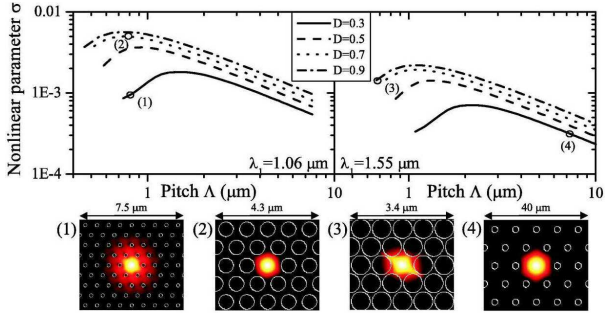}}
\caption{Double-log plots of $\sigma$ vs. $\Lambda$ for $\lf=10$~cm
  and $\tau=1$~ps. (1)-(4) show energy distributions of the FH modes
  from MPB. $n_{\rm SC}=9$ to ensure localized modes.}
\label{Fig:rho}
\end{figure}

The $\tilde \chi^{(2)}$ tensor is assumed to have the non-zero
elements $\tilde{\chi}^{(2)}_{jji}=\tilde{\chi}^{(2)}_{jij}=
\tilde{\chi}^{(2)}_{iij}=\tilde{\chi}^{(2)}_{iii}/3$, where $i$ is the
main direction of the poling voltage, and $j$ is either of the 2
remaining directions.\cite{kielich:1969} The present problem has two
degenerate solutions that are $x$- and $y$-polarized, respectively,
so it will suffice to consider $i=x$. We used a realistic value
$\tilde{\chi}^{(2)}_{xxx}=1$~pm/V. The nonlinear parameter $\sigma$ in
Fig.~\ref{Fig:rho} is calculated by fixing $\lambda_1$ and $D$ (in
contrast to Figs.~\ref{Fig:lambda-phase},\ref{Fig:lambda-1-155}, these
curves are not zero-GVM contours.) As expected $\sigma$ is seen to
become larger as $\Lambda$ is reduced as well as when $D$ is
increased, and we found the scaling $\sigma\propto
D/\Lambda=d/\Lambda^2$. $\sigma$ peaks when $\Lambda$ takes values
around the chosen fundamental wavelength $\lambda_1$, and drops for
$\Lambda<\lambda_1$ because the fundamental mode has maximum core
confinement at the peak [Fig.~\ref{Fig:rho}(2)], while it becomes more
poorly confined when $\Lambda<\lambda_1$ [Fig.~\ref{Fig:rho}(1,3)].
Instead, the SH having $\lambda_2=\lambda_1/2$ can stay confined
longer as $\Lambda$ is reduced, resulting in a poor modal overlap. [A
similar effect gives the cusp in $|\Delta n|$ in
Fig.~\ref{Fig:lambda-phase}(b).]  For large $D$, a decent fundamental
mode confinement is observed even for $\Lambda<\lambda_1$ [compare
Fig.~\ref{Fig:rho}(3) with (1),] giving a shift in the peak towards
smaller $\Lambda$. Finally, $\sigma$ gets larger when $\lambda_1$
becomes smaller because if $d$ and $\Lambda$ are fixed, the light is
better confined for smaller $\lambda$.  Table~\ref{tab:shg} shows
choices for efficient SHG with zero GVM using $\tau=1$~ps and
$\lf=10$~cm.  The large bandwidths imply that pulses as short as
$\tau_{\rm lim}=21$~fs can be converted.  The relative SHG
efficiencies $\eta/P_1$ are as high as 25 \%/mW.


\begin{table}[b]
  \centering
  \caption{Selected parameters for SHG with zero GVM. 
}   
  \label{tab:shg}
  \begin{tabular}{c|cc|ccc|cc}
$\lambda_1$ & $\Lambda$ & $D$ 
& $\bw$ & $\tau_{\rm lim}$ & $|\lcoh|$& $\sigma$ &
$\eta/P_1$\\ 
$\mu{\rm m}$ & $\mu{\rm m}$ &  
& ${\rm nm}$ &  fs& $\mu{\rm  m}$& $10^{-4}$& $\frac{\%}{\rm mW}$\\
\hline
0.80 & 0.70 & 0.96 &  13  & 73 & 2.1  & 112  & 25\\
1.06 & 0.85 & 0.72 &  77  & 21 & 3.7  & 49.8 & 6.3\\
1.55 & 1.60 & 0.43 &  170 & 21 & 14.4 & 11.5 & 0.5
  \end{tabular}
\end{table}

To conclude, by tuning the pitch and relative hole size in a standard
index-guiding silica PCF, we showed that SHG with zero GVM is possible
for any $\lambda_1>780$~nm.  This is a new and simple way to remove
GVM, which in addition has conversion bandwidths suitable for down to
20 fs pulse-conversion. The SHG nonlinear parameter was inversely
proportional to the pitch and proportional to the relative hole size,
due to smaller mode overlap areas for lower pitches and larger
relative hole sizes, and up to 25\%/mW conversion efficiencies was
found.

Support from The Danish Natural Science Research Council (FNU, grant no.
21-04-0506) is acknowledged. M. Bache's e-mail address is
bache@com.dtu.dk.


\end{document}